\documentclass[12pt]{article}
\usepackage{graphicx}
\def\gW{W}
\begin{document}

\title{Relativistic Model of Detonation Transition from Neutron to Strange Matter}

\author{ I.TOKAREVA\footnote{ Physics Department, Technion,
Technion-city, Haifa 32000, Israel; E-mail: iya@tx.technion.il },
A.NUSSER\footnotemark[1], V.GUROVICH\footnote{ Physics Institute
of NAS, Chui av. 265a, Bishkek, 720071, Kyrgyzstan }, and V.
FOLOMEEV\footnotemark[2]}
\date{}
 \maketitle


\begin{abstract}

We study the conversion of neutron matter into strange matter as a
detonation wave. The detonation is assumed to originate from a
central region in a spherically symmetric background of neutrons
with a varying radial density distribution.
We present  self-similar solutions for the propagation
of detonation in static and collapsing backgrounds of neutron matter.
The solutions are obtained in the framework of general relativistic hydrodynamics,
 and are relevant for the possible transition
of neutron into strange stars. Conditions for the formation of
either bare or crusted strange stars are discussed.

\end{abstract}

\section{Introduction}
Several authors have proposed that the ground state of matter is
in the form of bulk quark matter made of the three flavors,
up,  down and strange. We adopt the term strange matter (SM) to refer to
this bulk quark matter.
Conversion of nuclear matter to SM is suppressed
at ordinary nuclear densities. At very large nuclear densities
like those in neutron stars, where the Fermi energy is higher than
the strange quark mass, the conversion however is
possible.
This has led to conjecture that
 {\it strange stars} that are predominantly made of SM may form by conversion of   neutrons
in dense neutron stars~\cite{Bod,Witt}. The conversion is assumed
to be triggered at the center of the neutron star (where the
density is highest) either by
 presence of a seed SM or by spontaneous $\Lambda$ -hyperons.

Strange stars fall in a mass range similar to that of neutron
stars, but they have smaller radii\cite{Alcoc}. There is mounting,
albeit controversial, observational evidence for the existence of
strange stars. Likely candidates are the the compact objects 3C58
 and RX J1856.5-3754\cite{Helfand,Drake}.

 The dynamics of  transition into SM
 has not yet been  examined in details.
 According to ~
\cite{Alcoc,Olinto}, the transition is realized by outward
diffusion of strange quarks.  The increased fraction of SM as  a
result of diffusion triggers further conversion into SM. In this
case the conversion may be described as  slow combustion at speeds
of about $10^{7}{\rm m /s}$. However, stability analysis in the
presence of gravity  has shown that this combustion process is
unstable to small perturbations~\cite{HB}. Once these perturbation
grow they increase the effective area available for diffusion.
Therefore the conversion may actually propagate in the form of
detonation sustained by the enhanced ``burning''.
 In
 the simplest variant, the detonation wave (DW) is propagating into a neutron matter  NM  with
constant density. Newtonian gravity was used in ~\cite{Ben} to
determine the parameters of the  resultant strange star and to
estimate the energy
  released in the process.

 More
realistic models for the propagation of DWs should  take  into
account a varying  density of NM. Further, next to black holes,
neutron stars have the largest gravitational fields in nature.
Moreover, the characteristic time of gravitational processes
$t=(4\pi G\varrho)^{-1/2}$ is comparable to the hydrodynamical
time scale $t=R_{NS}/v_d\sim 10^{-5}$s\cite{Olinto}.
 Therefore, it is prudent
 to consider detonation in the context of
general relativistic hydrodynamics. In this paper we  study
self-similar models of DW propagation in non-uniform NM, taking
into account a self-consistent treatment of gravity according to
general relativity (GR). We consider detonation in static and
collapsing backgrounds of NM.

 The paper is organized as follow. In \S 2 we write the  equation of
 state for quark matter. In \S 3.1 we present the equations of special
 relativistic hydrodynamics in the absence  of  gravitation.
 Self-similar solutions in this case  are presented in \S 3.2
  for uniform  and  non-uniform
  backgrounds of NM.
 We turn to the general relativistic description in
 \S 4     where
we derive the equations for general  relativistic hydrodynamics
 and present the corresponding   numerical
 self-similar solutions for static and collapsing non-uniform background.
 In \S 5 we conclude with a discussion of the results and their implications.

\section{The Quark Matter Equation of State}

The equation of state (EOS) for hot (quark) matter has the form of
simple relation~\cite{Witt},
\begin{equation}
\label{eos3} p = {\frac{{1}}{{3}}}(\varepsilon - \varepsilon _{0}
), \quad \varepsilon _{0} = 4 B.
\end{equation}
We use throughout $\epsilon$ and $p$ to denote the energy density
and the pressure, respectively. This simple EOS may be obtained
straightforward from the next EOS~\cite{Ben}
\begin{equation} \label{eos1} p = {\frac{{19}}{{36}}}\pi ^{2}T^{4}
+ {\frac{{3}}{{2}}}T^{2}\mu ^{2} + {\frac{{3}}{{4\pi ^{2}}}}\mu
^{4} - B{\rm ;} \quad \tilde{n} = T^{2}\mu + {\frac{{1}}{{\pi
^{2}}}}\mu ^{3};
\end{equation}
\begin{equation}
\label{eos2} \varepsilon = {\frac{{19}}{{12}}}\pi ^{2}T^{4} +
{\frac{{9}}{{2}}}T^{2}\mu ^{2} + {\frac{{9}}{{4\pi ^{2}}}}\mu ^{4}
+ B; \quad s = {\frac{{19}}{{9}}}\pi ^{2}T^{3} + 3T\mu ^{2}.
\end{equation}
$T$ is the temperature, $\tilde{n}$ is the baryon number density,
$\mu$ is the chemical potential of quark matter, $B$ is the MIT
bag constant. The sound speed in such matter is defined as
$(\partial p/
\partial \varepsilon )_{s} = c_{s}^{2}$, where the derivation
occurs at constant entropy. It is easy to see that the sound speed
remains constant $c_{s} = 1 / \sqrt {3} $. The EOS \ref{eos3} was
used by Witten for calculation of the structure of strange
stars~\cite{Witt}. We shall adopt this simple approximation  for
modelling of detonation transition of neutron stars into strange
one.

\section{Detonation in Special Relativistic Hydrodynamics}

First, we describe the propagation of detonation without taking
into account the effects of gravity. This will be useful for
comparison of our results with the existing literature and will
also facilitate the interpretation of the general relativistic
solutions to be presented in \S\ref{GR}.
 The neglect of gravity is
justified if the detonation time scale is smaller than the
dynamical time scale. Gravitational effects become important after
the end of detonation, but in this section we will only be
concerned with describing the detonation phase.

\subsection{Basic equations}
Here we describe  the relevant equations for detonation in special
relativistic hydrodynamics (SRH). We assume that the  pressure of
NM is negligible compared to its
 energy density,
 \begin{equation}
\label{epn}
\varepsilon _{n}\gg p_{n},
\end{equation}
 otherwise there
will not  be any significant gain in pressure in the transition
from NM to SM. We will also assume  throughout the energy density
of the SM $\varepsilon_{s} $ behind the front is substantially
larger than the ''vacuum energy''  $\varepsilon_v=4B$ which is
close to nuclear energy density of usual matter,
\begin{equation}
\label{eps}\epsilon_{s}\gg 4B.
\end{equation}
 Under the conditions(\ref{epn})-(\ref{eps}) we express
these EOSs as
 \begin{equation}
 \label{en}
 p=n\varepsilon \; ,
 \end{equation}
where  $n=1/3$ and $n=0$ correspond to hot
 relativistic matter and cold neutron matter, respectively.
Values other than $n=0$ can be used to describe approximately the
EOS of the NM, in these cases $n$ can be regarded representing
 the square of an ``effective'' sound speed for the NM.

We will base  our description of  detonation in SRH on  an analogy
with  the classical theory of detonation~\cite{Land}. Therefore,
instead of using covariant notation we express the relevant
equations in terms of the radial component of the 3-velocity $v
(\tau, r)={\rm d } r/{\rm d}\tau$, where
 $r$ is  a radial coordinate and  $\tau=ct$ is the time measured  in units of lengths.
  We  use the notation
 $w = \varepsilon + p$ for the ``enthalpy''.

  The relevant equations of SRH are,
\begin{eqnarray}
\label{EoRH} {\frac{{1}}{{w}}}\left( {{\frac{{\partial
p}}{{\partial r}}} + v{\frac{{\partial p}}{{\partial \tau} }}}
\right) + {\frac{{1}}{{\gW ^{2}}}}\left( {{\frac{{\partial
v}}{{\partial \tau} }} + v{\frac{{\partial
v}}{{\partial r}}}} \right) = 0; \qquad \nonumber \\
{\frac{{1}}{{w}}}\left( {{\frac{{\partial \varepsilon} }{{\partial
\tau} }} + v{\frac{{\partial \varepsilon} }{{\partial r}}}}
\right) + {\frac{{1}}{{\gW ^{2}}}}\left( {{\frac{{\partial
v}}{{\partial r}}} + v{\frac{{\partial v}}{{\partial \tau} }}}
\right) + {\frac{{2v}}{{r}}} = 0,
\end{eqnarray}
where $\gW =\sqrt{ 1-v^{2}}$. \\
The first of these equations is the
Euler equation and the second is the continuity equation for
energy density.

These equations determine the evolution of the system in which a
detonation front propagates outward from the center of the neutron
star converting its into SM. In a frame of reference moving with
the front, a requirement of the energy - momentum flow be
conserved yields the following jump conditions at the front
\begin{eqnarray}
 (p_s + \varepsilon_s )v^f_s / \gW_s ^{2} = (p_n+\varepsilon
_{n}) v^f_{n} / \gW _{n}^{2};\qquad \quad\nonumber\\
\label{cond}\quad (p_s + \varepsilon_s )(v^f_s)^{2} / \gW_s ^{2} +
p_s = (p_{n}+\varepsilon _{n}) (v^f_{n})^{2} / \gW _{n}^{2}+p_n.
\end{eqnarray}
In addition, there is baryon flux conservation $n$
\begin{eqnarray}
{n_s}{v^f_s}/{\gW_s}={n_n}{v^f_{n}}/{\gW _{n}}. \end{eqnarray}
 Here the indexes  $``n", \, ``s"$ designate the
quantities concerning to NM and to SM, respectively. So, $v^f_n$
is the velocity of ``inflow" of neutron matter in front of DW and
$v^f_s$ is the departure velocity of SM from the front.

If the  3-velocity $v^f$ in (\ref{cond}) is measured  in the local
frame of detonation front then, according to the relativistic
equation of velocity transformation, in the laboratory local frame
it takes the value

\begin{equation}
\label{vel} v(D) = (D - v^f) / (1 - D \, v^f ),
\end{equation}
where $D$ is  the speed of DW in this frame.

The classical analysis of detonation conditions leads to
definition of a normal detonation condition when the speed of hot
fluid  corresponds to sound speed $v^f_s = c_{s}$~\cite{Land}. In
the case of NM at rest $$D = -v^f_{n},$$ and substituting
(\ref{en}) in the jump conditions (\ref{cond}) yields
\begin{equation}
\label{DW} v^f_s = c_{s}:\quad n=D(D-\sqrt{3}/2) /
(1-\sqrt{3}/2D), \quad
\varepsilon_{s}=D/(2/\sqrt{3}-D)\varepsilon_{n}.
\end{equation}

\subsection{Self-similar solutions in SRH}

Let the  detonation  wave propagate in cold NM background
($n\rightarrow 0$ according to(\ref{epn})), and let the  SM behind
the front be hot and relativistic (see (\ref{eps})). We will
obtain self-similar solutions of (\ref{EoRH}) for this situation.

 By
analogy to classical detonation, we find the self-similar solutions
 assuming the velocity $v$ to depend on $r$ and
$\tau$ only through the variable
$$
\xi = r / \tau.
$$

Analysis of the system (\ref{EoRH}) shows that self-similar
solutions can be  obtained in the case of power dependence of
density of NM,
$$
\varepsilon _{n} = \left( {r_{0} / r} \right)^{k}\,e_{n}, \quad
e_{n} = const.
$$
 This distribution of NM can be created by its own gravity
(Newtonian gravitation).
 But for a fast detonation wave (see
(\ref{DW})), the inhomogeneity  of  NM distribution
 is only important. (Similar situation is realized for a strong
shock wave in an inhomogeneous atmosphere~\cite{Raiz}.) According
to (\ref{DW}), we have at  the front position $r=r_{d}$,
\begin{equation}
\label{en_dist} \varepsilon_{s} = 3\,\varepsilon _{n} = 3\,\left(
{r_{0} / r_{d}} \right)^{k}\,e_{n},
\end{equation}
and the front velocity  remains constant. Therefore, the density
of SM at the front varies with time according to the relation:
$$
\varepsilon_{s} = 3\,(\tau _{0} / \tau )^{k}\,e_{n}.
$$
 This relation motivates
the following form for  $v$ and $\varepsilon_{s}$,
\begin{equation}
\label{sols} v = v(\xi ), \quad \varepsilon = (\tau _{0} / \tau
)^{k}\,e(\xi ).
\end{equation}

By expressing the partial derivatives $\partial /\partial r$ and
$\partial /\partial \tau$ in terms of  $d/ d \xi$ the Eqs.
(\ref{EoRH}) are transformed into the following ordinary
differential equations,
\begin{equation}
\label{EoRH2_1} {\frac{{dv}}{{d\xi} }} = {{{\left[ {{\frac{{2v\gW
^{2}}}{{\xi (1 - v\xi )}}} - {\frac{{3}}{{4}}}{\frac{{k\gW
^{4}}}{{(1 - v\xi )^{2}}}}} \right]}} \mathord{\left/ {\vphantom
{{{\left[ {{\frac{{2v\gW ^{2}}}{{\xi (1 - v\xi )}}} -
{\frac{{3}}{{4}}}{\frac{{k\gW ^{4}}}{{(1 - v\xi )^{2}}}}}
\right]}} {{\left[ {{\frac{{1}}{{c_{s}^{2}} }}\left( {{\frac{{v -
\xi} }{{1 - v\xi} }}} \right)^{2} - 1} \right]}}}} \right.
\kern-\nulldelimiterspace} {{\left[ {{\frac{{1}}{{c_{s}^{2}}
}}\left( {{\frac{{v - \xi} }{{1 - v\xi} }}} \right)^{2} - 1}
\right]}}},
\end{equation}
\begin{equation}
\label{EoRH2_2} {\frac{{d\ln e}}{{d\xi} }} = {\frac{{4}}{{(1 -
v\xi )}}}{\left\{ {{\frac{{(\xi - v)}}{{\gW
^{2}}}}{\frac{{dv}}{{d\xi} }} + {\frac{{vk}}{{4}}}} \right\}}.
\end{equation}

\subsubsection{Detonation Wave in Homogeneous Background of NM}
We first present solutions for the  propagation of DW in  a
homogeneous distribution of NM ($k=0$). In the case the equations Eqs.(\ref{EoRH2_1})-(\ref{EoRH2_2}) become
equations is
\begin{equation}
\label{EoRH1} {\frac{{dv}}{{d\xi} }} = {{{\left[ {{\frac{{2v\gW
^{2}}}{{\xi (1 - v\xi )}}}} \right]}} \mathord{\left/ {\vphantom
{{{\left[ {{\frac{{2v\gW ^{2}}}{{\xi (1 - v\xi )}}}} \right]}}
{{\left[ {{\frac{{1}}{{c_{s}^{2} }}}\left( {{\frac{{v - \xi} }{{1
- v\xi} }}} \right)^{2} - 1} \right]}}}} \right.
\kern-\nulldelimiterspace} {{\left[ {{\frac{{1}}{{c_{s}^{2}
}}}\left( {{\frac{{v - \xi} }{{1 - v\xi} }}} \right)^{2} - 1}
\right]}}}; \quad {\frac{{d\ln \varepsilon }}{{d\xi} }} =
{\frac{{4}}{{{\gW ^{2}}}}}\left({{\frac{{\xi - v}}{1 - v\xi
}}}\right){\frac{{dv}}{{d\xi} }},
\end{equation}
 and in  the non-relativistic case
($v, c_{s} < < 1$) the equation for $v$ reduces to
Zeldovich's classical equation for detonation~\cite{Land},
$$
{\frac{{dv}}{{d\xi} }} = {\frac{{2v}}{{\xi} }} {\frac{{c_{s}^{2}
}}{{\left( {\left( {v - \xi}  \right)^{2} - c_{s}^{2}} \right)}}}.
$$

\begin{figure}[h]
 \begin{center}
  \includegraphics[height=6 cm,]{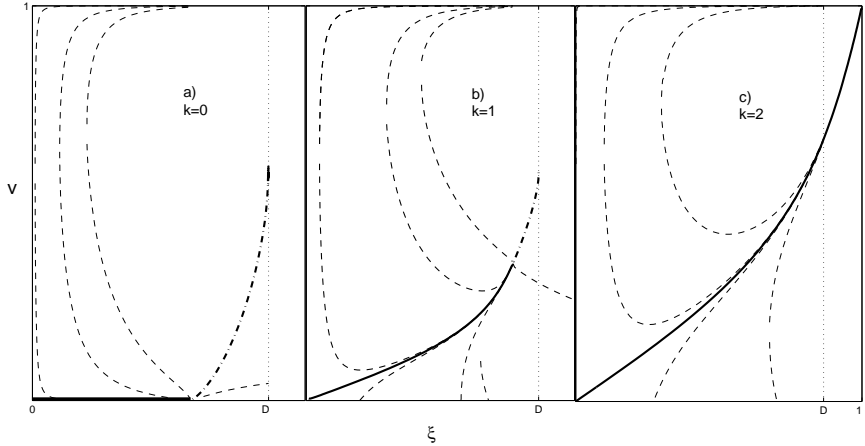}  \caption{A phase plane of the solutions  of
  (\ref{EoRH2_1})
   for propagation of detonation wave in
     a homogeneous  (a) and an inhomogeneous background
     (b,c). In (a) there is a solution for velocity $v(\xi)$ of gas that
       leaves detonation
      front  propagating  with  velocity $D$
       and  comes to  node of solutions $\xi=1/\sqrt{3}, v(\xi)=0$,
      where it may be matched to the solution for rest matter
      $v=0$.  For $k=1$ (panel b)  node shifts in the plan $(\xi, v)$.
       For $k=2$ (panel c)
  there in no node but we have a unique solution  leaving detonation front and converging  to $v=0$}
  \label{SR1}\end{center}
\end{figure}

Analysis of the  first equation of (\ref{EoRH1}) in the phase
plane $(\xi,v)$ reveals a variety of solutions which all converge
to a node at $\xi=c_{s}=1/\sqrt{3}$ and $v=0$ with  the derivative
$d v/d\xi=0$. The solutions can be obtained independently for the
ranges $0\le \xi<c_{s}$ and the $c_{s}\le \xi<D$.  The solutions
can be found by numerical integration of the equation
(\ref{EoRH1}) starting from an arbitrary point in phase space.

Examples of these  solutions  are shown in figure (\ref{SR1}, a).
As is clear from this figure the only physically accepted solution
in the range $0\le \xi <c_{s}$ is $v(\xi)=0$ (see~\cite{Land})
shown as the thick solid line. To the right of $\xi=c_{s}$ the
solution shown as the dashed dotted line  corresponds to normal
detonation. For this solution the  velocity at the front $\xi=D$
is equal to the sound speed (see (\ref{DW})) and is zero at
$\xi=c_{s}$. The derivative of the velocity diverges as $\xi $
approaches $D$,
 and tends to zero at $\xi \to c_{s}$.
This detonation solution describes hot  matter at rest for
$\xi<c_{s}$ and moves outward for $c_{s}<\xi<D$. The velocity of
the front  is $\sqrt{3}/2$, according to
(\ref{DW}).

  Our solution for the propagation of detonation in a uniform distribution
  of NM
  is fully consistent with previous calculations in the
  literature~\cite{Ben}.
We also refer the reader to the literature treating the similar
problem for relativistic detonation in ``scalar field''
stars~\cite{Fol}.

\subsubsection{Detonation Wave  in an Inhomogeneous
Background of NM} \label{ISR}

 We search now for self-similar
solutions of Eq. (\ref{EoRH2_1}) for $k>0$, i.e., for the
propagation of detonation in NM with decreasing density. For
$k>0$, also,  the equation has a node to which all solutions
converge with the identical derivative. In this case, however, the
node is at $v>0$ and not at $v=0$ as in the $k=0$ case. Examples
of the solutions are shown in panels b and c of Fig.1.
 As in the homogeneous case,
the physical solution corresponds to   with $v = 0$ at $\xi=0$.
The point $\xi=v=0$ is a  critical one (saddle) and   the solution
originating from this point is the   separatrix and is shown as
solid line in the figure.  Near  $\xi = v = 0$ we find that
\begin{equation}
\label{separ} v = k\,\xi / 4 \; ,
\end{equation}
for $\xi \ll 1$.
 For $k<4/3$ the node lies to the left of DW wave
at $ \xi<D$. Therefore, self-similar solutions corresponding to
normal detonation (cf. \ref{DW})) exist only for $k\le4/3$.

  For
$k> 4 / 3$ the node shifts  to the right of the DW. In this case,
there are no self-similar solutions for normal detonation in a
static neutron configuration. But we can find self-similar
solutions for over-compressed detonation. For such type of
detonation the velocity of the hot matter at the front is
determined by the details of the actual physical situation but it
is less than sound speed in SM~\cite{Land} The position of DW
front is defined from (\ref{cond}), $v^f = -v_{d} $,
\begin{equation}
\label{placeDF}v_d<c_s:\quad D = {\frac{{\varepsilon_{s} \,v_d^{2}
+ p_{s}}}{{(\varepsilon_{s} + p_{s})\,v_d}}};\quad w_{s} =(n+1)
\varepsilon _{n} \left( {{\frac{{1 - v_d^{2}}}{{1 - D^{2}}}}}
\right)\,{\frac{{D}}{{v_d}}} = \Gamma (D)\,\varepsilon _{n}.
\end{equation}
Then a laboratory velocity of outgoing SM from the front of wave
(\ref{vel}) will be supersonic nearby the front of DW but
transition through the sound speed in the self-similar solution
occurs continuously and $v \to 0$ at $\xi \to 0$ according to the
formula (\ref{separ}).

The conditions (\ref{placeDF}) are actually conditions of
so-called ``over-compressed'' detonation.

 In our statement of the
problem, at NM densities less than $10^{16}$ $\mathrm{g/cm^3}$ we
deal with a transition mechanism based on diffusion of strange
quarks from SM to NM. Corresponding velocity of diffusion $v_d$
depends on density of NM~\cite{Olinto}, but in principle, it
should be less than sound speed in SM.
  Therefore the
velocity of SM leaving the front is equal to the  velocity of the
anomalous (turbulent) diffusion $v_d$, which may be higher than
the estimate given in \cite{Olinto}).

\section{Detonation in General Relativistic  Hydrodynamics}
\label{GR}

\subsection{Basic Equations}

As we shall consider a
spherically-symmetric problem, it is convenient to write the
metric as
\begin{equation}
\label{metrics} ds^{2} = e^{\nu} d\tau^{2} - e^{\lambda} dr^{2} -
r^{2}d\Omega ^{2}; \quad d\Omega ^{2} = d\theta ^{2} + \sin
^{2}\theta \,d\varphi ^{2}.
\end{equation}
Here the standard spherical system of coordinates $r, \theta ,
\varphi $ is used. Variables $\nu, \lambda $ are functions of $r,
\tau$.

To derive self-similar solution in form similar to SRH analysis
(\ref{EoRH}) we will use an ``auxiliary'' time $\eta$ instead of
the proper time of local observer $\tau$. The transformation from
variables $\tau, r$ to new variables $\eta, r$ is carried out with
help of the Jacobian\cite{Shar}:
\begin{equation}
\label{jacob} \frac{\partial (t; r)}{\partial (\eta;
r)}=\frac{\partial t}{\partial \eta}= \exp\left({\frac{\lambda
-\nu}{2}}\right).
\end{equation}
The system of hydrodynamics equations  in GR are then
\begin{equation} \label{EoRHGR1_1} {\frac{{1}}{{w}}}\left(
{{\frac{{\partial p}}{{\partial r}}} + v{\frac{{\partial
p}}{{\partial \eta} }}} \right) + {\frac{{1}}{{\gW ^{2}}}}\left(
{{\frac{{\partial v}}{{\partial \eta} }} + v{\frac{{\partial
v}}{{\partial r}}}} \right) = - {\frac{{1}}{{2}}}\left(
{{\frac{{\partial \nu} }{{\partial r}}} + v{\frac{{\partial
\lambda} }{{\partial \eta} }}} \right);
\end{equation}

\begin{equation}
\label{EoRHGR1_2} \left( {{\frac{{\partial \ln
\varepsilon}}{{\partial \eta} }} + v{\frac{{\partial \ln
\varepsilon}}{{\partial r}}}} \right) + {\frac{{1}}{{\gW
^{2}}}}\left( {{\frac{{\partial v}}{{\partial r}}} +
v{\frac{{\partial v}}{{\partial \eta }}}} \right) +
{\frac{{2v}}{{r}}} = - {\frac{{1}}{{2}}}\left( {{\frac{{\partial
\lambda} }{{\partial \eta} }} + v{\frac{{\partial \nu }}{{\partial
r}}}} \right).
\end{equation}
The Einstein equations for the metrics can be written down as:
\begin{eqnarray}
\label{EoGR} {\frac{{\partial \lambda} }{{\partial \eta} }}\left(
{1 + v^{2}} \right) + v\left( {{\frac{{\partial \nu} }{{\partial
\,r}}} + {\frac{{\partial \lambda }}{{\partial \,r}}}} \right) =
0, \quad {\frac{{\partial \lambda }}{{\partial \,\eta} }} = -
{\frac{{\kappa \,(p + \varepsilon )r\,v\,e^{\lambda }}}{{\gW
^{2}}}},\nonumber.\\
 {\frac{{\partial \lambda} }{{\partial \,\eta}
}} + v{\frac{{\partial \lambda }}{{\partial \,r}}} = - v\left(
{{\frac{{e^{\lambda}  - 1}}{{r}}} + \kappa \,p\,r\,e^{\lambda} }
\right),\qquad \qquad
\end{eqnarray}
where $G,c$ are Newtonian gravitation constant and speed of light,
respectively and $\kappa = 8\pi G / c^{4}$. The equations
(\ref{EoRHGR1_1})-(\ref{EoGR}) can be presented as system of three
equations by eliminating of function $\nu (r,\,\eta )$,
\begin{eqnarray}
\label{fullS} {\frac{{1}}{{p+\varepsilon}}}\left(
{{\frac{{\partial p}}{{\partial r}}} + v{\frac{{\partial
p}}{{\partial \eta} }}} \right) + {\frac{{1}}{{\gW ^{2}}}}\left(
{{\frac{{\partial v}}{{\partial \eta} }} + v{\frac{{\partial
v}}{{\partial r}}}} \right) = {\frac{{1}}{{2v}}}\left(
{{\frac{{\partial \lambda} }{{\partial \eta} }} +
v{\frac{{\partial \lambda} }{{\partial r}}}}
\right), \nonumber \\
{\frac{{1}}{{p+\varepsilon}}}\left( {{\frac{{\partial \varepsilon}
}{{\partial \eta} }} + v{\frac{{\partial \varepsilon} }{{\partial
r}}}} \right) + {\frac{{1}}{{\gW ^{2}}}}\left( {{\frac{{\partial
v}}{{\partial r}}} + v{\frac{{\partial v}}{{\partial \eta} }}}
\right) + {\frac{{2v}}{{r}}} = {\frac{{v}}{{2}}}\left(
{{\frac{{\partial \lambda} }{{\partial r}}} +
v{\frac{{\partial \lambda} }{{\partial \eta} }}} \right), \nonumber \\
{\frac{{\partial \lambda} }{{\partial \,\eta} }} = -
{\frac{{\kappa \,(p + \varepsilon )r\, v\,e^{\lambda}}}{{\gW
^{2}}}}.\qquad\qquad\qquad
\end{eqnarray}
In the limit of $\lambda=0$ this system of equations reduces to the
SRH equations  (\ref{EoRH2_2}).

\subsection{Self-similar solutions in GRH}

 In
contrast to SRH, where self-similar solutions could be found for
all $k$, in GRH  the full system (\ref{fullS})  permits only
self-similar solutions only with  NM distribution density
(\ref{en_dist}) with $k = 2$. We show in appendix \ref{ap} that
this profile describes  approximately a realistic density profile
in a neutron star with a Harrison-Wheeler-Wakano equation of
state\cite{Haris}. In analogy with SRH we make the ansatz,
\begin{equation}
\label{approx2} \xi = r / \eta,\quad \kappa p\,(\eta ,r)
=2\mathit{\Pi} (\xi)\,/\eta ^{2}, \quad v\,(\eta ,r)=v(\xi),\quad
\lambda=\lambda(\xi)
\end{equation}
Adopting an EOS of the form (\ref{en}) and using
 $v,\, \mathit{\Pi},\, \lambda$ expressed in terms of the self-similar
 variable $\xi$, the GRH equations are  expressed as,
\begin{eqnarray}
\label{full1}  {\frac{{dv}}{{d\xi} }} =\left[{\frac{{2nv\gW
^{2}}}{{\xi (1 - v\xi )}}} - {\frac{2n{\gW^{4}}}{{(n+1)(1 - v\xi
)^{2}}}}  + \frac{(n+1)\mathit{\Pi} e^{\lambda}}{n}\left\{ f_1^{2}
- {n}{v^{2}}\right\}\right]/f_2
\end{eqnarray}
\begin{eqnarray}
\label{full2}{\frac{{d \mathit{\ln\Pi}}}{{d\xi} }} =(n+1) {\left[
{{\frac{f_1}{{(1 - v\xi )}}}\left( {{\frac{{2}}{{n+1}}} -
{\frac{{2v}}{{\xi} }}}\right)-{\frac{{v}}{{2{(1 - v\xi )}}}}  -
\frac{(n+1)\mathit{\Pi} e^{\lambda}}{n}  f_1} \right]}/f_2
\end{eqnarray}

\begin{equation}
\label{full3} {\frac{{d\lambda} }{{d\xi} }} =
{\frac{{2\,(n+1)\mathit{\Pi} e^{\lambda} v}}{{n\,\gW ^{2}}}},
\qquad f_1={{\frac{v - \xi}{1 - v\xi}}}, \qquad f_2={f_1^{2} -n }.
\end{equation}

In the case of hot SM ($n=1/3$) , Eqs. (\ref{full1})-(\ref{full3})
have a form,
$$
{\frac{{dv}}{{d\xi} }} = {\left[ {{{ {{\frac{{2v\gW ^{2}}}{{\xi (1
- v\xi )}}} - {\frac{3{\gW ^{4}}}{{2(1 - v\xi )^{2}}}}} }} +
\frac{ \Pi e^{\lambda}}{2} {\left\{3 \left({\frac{v - \xi}{1 -
v\xi}}\right)^{2} - {{v^{2}}} \right\}}}
\right]}/\left[3\left({\frac{v - \xi}{1 -
v\xi}}\right)^{2}-1\right], $$
\begin{equation}
 {\frac{{d \ln \Pi}}{{d\xi} }} =
{\frac{{4}}{{(1 - v\xi )}}}{\left\{ {{\frac{{(\xi -
v)}}{{\gW^{2}}}}{\frac{{dv}}{{d\xi}
}}+{\frac{{v}}{{2}}}+\frac{{(v-\xi)}}{{2v}}}{\frac{{d\lambda}}{{d\xi}
}}\right\}},\qquad \qquad  \label{fd3} {\frac{{d\lambda} }{{d\xi}
}} = {\frac{{\Pi e^{\lambda} v}}{{\gW ^{2}}}},
\end{equation}
where $\Pi = 8\mathit{\Pi}$.

 We first investigate the  boundary conditions imposed on this system of
 equations~ (\ref{fd3}). We
assume that at  $\xi = 0$ we have
\begin{equation}
\label{init} v(0) = 0, \quad \Pi (0),  \lambda (0) = const.
\end{equation}
 In the vicinity of $\xi=0$ we
 solution can be approximate, up to third order in $\xi$, as
\begin{equation}
\label{sol0} v = \alpha \xi , \quad \Pi = \Pi _{0} - \beta \xi
^{2}, \quad e^{ - \lambda} = 1 + \gamma \xi ^{2},
\end{equation}
where
$$
\alpha = 1 / 2, \quad \beta = \Pi _{0} (\Pi _{0} / 2 - 1), \quad
\gamma = - \Pi _{0} / 4.
$$

\subsubsection{Detonation in a static background of NM}
The  static   solution  of the system (\ref{full3}-\ref{full1})
for cold  NM background
 \begin{equation}
 \label{static}
 v=0,\quad \mathit{\Pi}=C_1/\xi^2\quad
 e^{-{\lambda}}=C_2.
 \end{equation}
is also the analytical solution of the Oppenheimer-Volkoff
equation (see Appendix)in the case $n\ne0,
\;C_2=C_1((n+1)/2n)^2$.

\begin{figure}[tb]
\centering{\includegraphics[width=100mm]{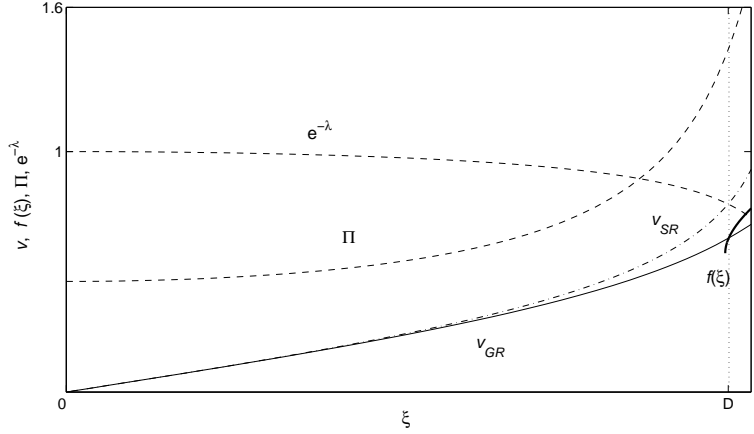}} \caption{The
dynamical variables  $\Pi(\xi)$, $v_{GR}(\xi)$ and
$e^{-\lambda(\xi)}$ obtained from  the   self-similar solution of
Eqs. (\ref{fd3}). There is a small difference between the velocity
in $v_{GR}$ and $v_{SR}$ obtained, respectively,
 from the general and special relativistic solutions.  }
 \label{fig3}
\end{figure}

 This full solution of \ref{fd3} for hot quark matter is found
  numerically and  shown as the solid line in Fig.\ref{fig3}.
 It corresponds to the separatrix
  by analogy to the same problem in
SRH (shown as the dash-dot line in the figure) and is valid
mathematically up to some $\xi<1$. However, the solution must be
terminated at the DW position, $D$.
In the case of normal detonation, the  the position $D$ on the
axis $\xi$ is found  by the requirement that
 the speed of outgoing SM $v^f$ is equal to the sound
speed.
The previous analysis has shown, however,  that for
 DW propagating  in a  background $\rho _{n} \sim 1
/ r^{2}$ such point can not exist  without taking into account
gravitation. The gravitational influence results in  backward
motion of node towards the front of normal detonation in the phase
plane but jump conditions of normal detonation exist  only in hot
neutron stars with $n\rightarrow1/3$.  This  dynamical process may
convert hot NS into SS without any appreciable  thermal energy
release.
 In cold NS we must deal with over-compressed detonation (see
section \ref{ISR}). A numerical calculation shows that $v^f$ can
be  a little less than $c_{s}$.

For subsequent analysis of numerical models of transition
 we will be  using analytical approximation  of neutron
star (see \ref{ap}). It satisfactorily describes distributions of
density $\delta$, pressure $\Pi$ and the metrics $\lambda$ of
neutron star from some initial dimensionless radius  $x_{0}$. Thus
\begin{eqnarray}
\label{delt} \Pi =8n A(n) / x^{2}, \quad A(n) = 2 n / ( n^{2} + 6
n + 1),\\
 \label{mtr}
 - g^{11} = e^{ - \lambda}  = 1 - 2A = const.\qquad
\end{eqnarray}

According to (\ref{placeDF}), the pressure of SM matter behind the
DW front with account of (\ref{approx2}) will be defined by
relation:
\begin{equation}
\label{rel} 4\,p(\eta ,r) = 4\,\mathit{\Pi}(\xi ) / \eta ^{2} =
(n+1)\Gamma (D)\,A(n) / x^{2},
\end{equation}
since the front of wave  propagates with constant speed $D$ and
 time $\eta$, as well as $r$, is measured in units of $L_{\ast}$
and $\eta = x/D$. The initial moment of time is chosen so that the
front of wave   is found at some $x_{0}$ where the analytical
approximation of distribution of  NS density is true. (Note, that
true distribution has a maximum in the center and a singular
analytical solution approximates only the decreasing of this
density with growth of $x$ starting with finite $x_0$. It is
supposed that the specified self-similar solution realizes from
the moment of achievement of point $x_{0}$  by the DW front and
further.) Thus for the representative of pressure at the front of
wave  we have
\begin{equation}
\label{pres} \Pi(D) = 2 \Gamma (D) (n+1)A(n) /  D^{2}.
\end{equation}

And position of $D$ is defined in the following way: at the point
$\xi=D$ values $\xi$, $v$ and $v^f$ have to obey the law of
energy-momentum flow conservation (\ref{cond}) and the
relativistic equation of velocity transformation simultaneously:
\begin{equation}
\label{defd}
 \xi=\frac{(3(v^f)^2+1)(n+1)}{8(v^f)}+\frac{\sqrt{n^2-10n/3+1}}{2},
 \qquad v=\frac{\xi+(v^f)}{1+\xi (v^f)}=
 f(\xi)
 \end {equation}
Thus, we are starting the numerical solution   from point $\xi_0 =
0$ taking into consideration an analytical research of the
solution in the vicinity of $\xi_0 $ (\ref{init}). At that, the
setting of $\Pi_0$ and $\lambda_0$ should be chosen so that after
definition of the front position they would be  corresponding to
the mentioned values (\ref{pres}) and (\ref{mtr}).

\subsubsection{Detonation in a collapsing background of neutron
matter}
\label{collapse} Recent observational data strongly
suggest that
 ``long-duration'' Gamma Ray Bursts (GRBs) are associated with supernova
explosions\cite{Bomb}. These GRBs follow  the supernova events
after a period varying from hours to months. This may point to the
possibility that these GRBs appear as a result of the conversion
of  NM into SM during collapse of the SN remnant. This occurs when
 the mass of the NM  becomes larger than the Chandrasekhar
mass limit. The detonation in this case propagates on the
background of the falling NM, so normal detonation wave forms.

Negligible  pressure forces   in the collapsing NM allow us to
neglect  pressure in this statement of the problem. There are only
two relevant parameters determining the self-similar solutions.
These are the mass and the radius  of the remnant strange star.
The parameters of obtained self-similar solution are compare with
those of the  observational data.
\begin{figure}[h]
\centering{\includegraphics[width=80 mm,height=50 mm]{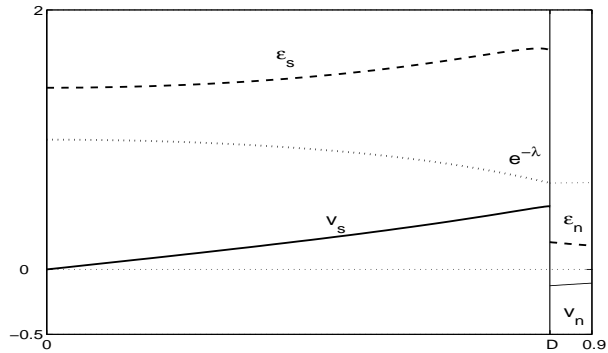}}
\caption{ The dynamical variables obtained in the self-similar
solutions in a collapsing background of NM with $n=0$.}
\label{col}
\end{figure}

In this statement of problem the jump conditions on the front
(\ref{cond}) yield
\begin{equation}
\label{con_dyn} v_s^f=1/\sqrt{3},\; v_n^f=\sqrt{3}/2, \;
\varepsilon_s=3\varepsilon_n.
\end{equation}

As in the stationary case, we are solving system (\ref{fd3}) for
hot strange matter  from point $\xi_0$ to place of the front
\begin{equation}
\label{front_dyn} D=\frac{v-v_n^f}{1-v\, v_n^f},
\end{equation}
The solutions of system (\ref{full1})-(\ref{full3}) for NM and SM
are presented in Fig.(\ref{col}). At that we will claim the metric
$e^{-\lambda}=1-2MG/Rc^2$ to be defined by mass $M$ of the strange
star at the radius $R$ (at $r=R \quad\rho c^2=4B$). Here we
use parameters $M=1.1 M_{\odot}$ and $R=9.7 km$
 (see data in \cite{Bomb} ).

\subsection{Numerical estimates}
\label{ne} As an example of detonation in static neutron star, let
us consider the model of detonation transition of neutron star
with mass of central part $M\sim 0.59 M_\odot$ and radius $\sim
8.4 \,\mathrm{km}$ measured in characteristic lengths $L_{\ast} =
2.8 \cdot 10^{5}\, \mathrm{cm}$ that corresponds to $x =1.4$. The
analytical approximation of density distribution $\rho _{n} $
measured in units of $ \rho _{\ast}  = 1.34 \cdot 10^{16}
\mathrm{g/cm^3}$ is set by expression (\ref{delt}) with $n = 1 /
12$. At that $A = 0.111, \, e^{ - \lambda}  = 0.76$. The results
of numerical research for the over-compressed detonation are
submitted in Fig.~\ref{fig3}.

In the example under consideration the velocity of DW  is equal
$\sim $0.85, that is close to the velocity $D =\sqrt {3} / 2 =
0.851$.

 It is interesting to estimate the value  energy concerned
with propagation of hot matter behind  DW  front, when the last
one reaches the specified boundary radius of neutron configuration
$x = 1.4$:
\begin{equation}
\label{kin1} E= 4\pi {\int\limits_{0}^{R} {r^{2}{\left[ {T_{0}^{0}
- T_{0}^{0} (v = 0)} \right]}}} dr = 16\pi {\int\limits_{0}^{R}
{{\frac{{v^{2}}}{{\gW ^{2}}}}p\,r^{2}}} dr
\end{equation}
whence, using (\ref{rel}) and (\ref{pres}), we shall have
\begin{equation}
\label{kin2} E = {\frac{{4\pi xL_{\ast}} }{{\kappa Dc^2}}}
{\int\limits_{0}^{D} {{\frac{{v^{2}}}{{\gW ^{2}}}}\Pi \,\xi ^{2}}}
\,d\xi.
\end{equation}
The value of the dimensionless integral, obtained on the basis of
the numerical solution, gives the value $0.057$, corresponding to
$E = 4.8 \cdot 10^{52} \mathrm{erg}=0.027 M_{\odot}$. This is
close to the the  energy produced in a  gamma-ray
burst~\cite{GRB}.

The gravitational mass defect  $\Delta
m$. According to~\cite{Land1}  we have,
$$
\Delta \,m = 4\pi {\int\limits_{0}^{R} {T_{0}^{0} \left(
{e^{\lambda / 2} - 1} \right)r^{2}d\,r}}  = 4\pi
{\int\limits_{0}^{R} {p\left( {{\frac{{4}}{{\gW ^{2}}}} - 1}
\right)\left( {e^{\lambda / 2} - 1} \right)r^{2}d\,r}}
$$
or, using (\ref{rel}) and (\ref{pres}),
$$
\Delta \,m = {\frac{{4\pi x   L_{\ast} }}{{\kappa D c^2
}}}{\int\limits_{0}^{D} {\Pi \left( {1 / \gW ^{2} - 1 / 4}
\right)\,\left( {e^{\lambda / 2} - 1} \right)\,\xi ^{2}d\,\xi} }
{\rm .}
$$
   Thus,we have
$\Delta m = 1.3 \cdot 10^{52} \mathrm{g}= 0.007M_{\odot}$.

The energy release associated with detonation can be roughly estimated
as $E-\Delta m$.  In our model it is only about $4\%$ of
the mass  $M$.

The detonation will stop when the NM density is close to
$4B/c^{2}$, leaving a crust made of the original matter of
the neutron star.
This crust can be expelled if  the kinetic energy of the SM
is large enough.
The kinetic energy needed for that can be estimated by
comparing the binding energy of the crust,
$$E_b=4\pi
{\int\limits_{R}^{R_{\ast}}\left( {\rho e^{\lambda / 2} -
\rho}\right) r^{2}d\,r} $$ to the energy release, $E-\Delta M$. We
can see that in the case $n=1/12$, $ E_b=0.005M_{\odot}$ which is
less than the energy release $E-\Delta m=0.013M_{\odot}$.
 Therefore for this value $n$ the crust might be expelled, leaving a bare strange star.
Table 1 lists estimates for other values of $n$ and for collapsing
NM.
\begin{table}[htbp]
 \caption{ \small The parameters of detonation in static ($v_n=0$) and
collapsing ($v_n=-0.13$) background of neutron matter. Here
$\rho_c $ is the central neutron density prior to detonation
 (in $10^{15}$~g/cm$^{3}$),  $M$ - mass of  SM core,  $R$
 is the radius at which detonation stops (in km), $v_s$ -speed of
 SM behind the detonation front,
 $E$ is the SM  ``kinetic energy'',
 $\Delta m$ is $E_b$ the  binding energy of SM core and NM crust,
  correspondingly.   All energies are in $10^{51}$ erg.}
\begin{center}
{\begin{tabular}{ccccccccc}
\multicolumn{8}{c}{} \\[10pt]\hline
$\rho_c $ & n & $M/M_{\odot}$ & $R$ &  $v_s$&$E$& $\Delta m$ & $E_b$ \\
\hline\hline 9.40&0.2&0.59& 5.23&-0.51&35.18& 38.49&20.94\\
6.71&0.16& 0.60& 5.42& - 0.52 &50.00&21.35& 22.13\\
5.37&0.08 &0.573&8.37&-0.52&48.44& 13.24&28.27\\
\hline \hline
collapse&0&1.1&9.74&-1/$\sqrt{3}$& 87.72& 58.64&-\\
 \hline
\end{tabular}}
\end{center}
\end{table}

\section {Discussion}

We have studied the conversion of neutron matter into strange
matter as a detonation wave propagating outward of a central
region. The dynamics of neutron-quark transition has been studied
earlier  for a uniform static background of neutron matter in the
context of special relativity~\cite{Olinto}. Here we employ
special and general relativistic hydrodynamics  to find
self-similar solutions for the propagation of detonation in a
non-uniform background of neutron matter. Both static and
collapsing backgrounds are considered. Solutions for detonation in
a static background may correspond to the conversion of neutrons
in an otherwise  stable neutron star.
 In a collapsing background the solutions may describe
 conversion in the neutron core of SN Type II~\cite{Bomb,ShapT}.
In this latter case detonation is assumed to  be
 triggered at the onset of the
 collapse.

A description of the physical processes leading to detonation can
be found in papers~\cite{Olinto,HB}. Briefly, the initial stages
of conversion of neutrons to strange quark matter proceeds by
diffusion of SM into the neutron matter. The diffusing SM then
stimulates further conversion and the whole process may be
described as combustion~\cite{Olinto}. However, the process turned
out to be hydrodynamically unstable when gravity is taken into
account~\cite{Lug}. Hydrodynamical instabilities increase the
surface area available for diffusion allowing for the possibility
of propagation in the form of detonation. In the other words,
these processes leads to anomalous diffusion of strange quarks to
the detonation front  with speed that is equal to speed of leaving
hot matter.

In special relativity the self-similar solutions describe
detonation in a static  background of  NM with distribution $\rho
\sim r^{-k}$.
 It was shown  the condition of a normal detonation exist
  for  $k<4/3$, while
  denotation must be over-compressed for $k\ge 4/3$.

Self-similar solutions of the problem in general relativity are
permitted only for  $k=2$. The reason for this restriction is
 the appearance of  new fundamental constant $G$ in GR.
  This additional physical scale prevents the
development of a self-similar behavior of the type considered
here. In  special relativity the detonation for $k=2$ is
over-compressed independently of the equation of state of cold
matter. However, in general relativity we obtain normal detonation
for $n$ close to $1/3$ ($P=n\varepsilon$). Below a limiting value
of  $n$ only over-compressed detonation is possible. We have  not
been able to find the limiting value of $n$ analytically but  our
numerical calculations show that it should be $n\approx 0.22$.

Normal detonation is obtained for  conversion in collapsing cold NM.
 In this case the  pressure of NM is completely neglected, i.e.,
 $n=0$ in (\ref{en}). In
such statement of problem
one can   easily calculate all relevant
quantities using only two parameters characterizing the system, e.g.,
the  mass and
radius  of the strange star.

 The
propagation of detonation in neutron stars must halt at the radius
where the neutron density is close to the quark vacuum energy,
$\rho=4B/c$. When the DW reaches this radius the SM would be
surrounded by a neutron crust.
The description in this paper breaks down at this
stage of the evolution.
However,
we still can use the self-similar solutions to
determine the energetics of the system and
to predict some general features its subsequent evolution.
In all of our self-similar solutions, except those
describing detonation in a uniform background,  the velocity of SM
is moving outward. Although the ``kinetic energy'' of the SM
 is insufficient to disassemble the whole configuration (see Table 1),
 it might be enough to expel the crust surrounding the quark core. According to
  Table 1, if  $n$ is close to $1/3$, the crust is maintained. In this case we
may end up with a crusted strange star. For lower $n$ the energy
is sufficient to remove the crust, leaving a  a bare strange
star.
 In our estimates presented in Table 1
  we      have neglected a few factors.
Non-vanishing mass of strange quark and constant of strong
interaction taken into account will lead to reducing of energy
release.
  We also have neglected any further conversion
(e.g., by slow combustion in the crust) after the detonation is
halted.

\section*{Acknowledgements}

We wish to thank A.Starobinsky for useful discussion of the model
of relativistic detonation which have been considered by V.G. and
V.F.  in~\cite{Fol}. Also we acknowledge to V.Usov  for
affirmative remarks that they made.

\section{Appendix A: Density profile of  Neutron stars with a HWW equation of state}
 \label{ap}
 In this section we solve  the Oppenheimer-Volkoff equation
 to compute the density profile in neutron stars
with the Harrison-Wheeler-Wakano (HWW) equation of state
 \begin{equation}
\label{HW} \Pi = {\frac{{1}}{{3}}}\left( {{\frac{{\delta ^{5 /
3}}}{{(1 + \delta ^{5 / 9})^{6 / 5}}}}} \right),
\end{equation}
 where $\Pi = p /\rho _{\ast} c^{2}, \quad  \delta = \varepsilon/ \rho _{\ast}c^{2}, \quad \rho
_{\ast}  = 1.34 \cdot 10^{16} \mathrm{g/cm^3}$. The HWW EOS is
valid for central densities $>4.63\cdot 10^{12} \mathrm{g/cm^3}$
~\cite{Haris}. The sound speed $c_{s} $  is given by
\begin{equation}
\label{snd} \left( {{\frac{{c_{s}} }{{c}}}} \right)^{2} =
{\frac{{dp}}{{d\varepsilon} }} = {\frac{{d\Pi} }{{d\rho} }} =
{\frac{{1}}{{9}}}  {\frac{{\delta ^{2 / 3}(5 + 3\delta ^{5 /
9})}}{{(1 + \delta ^{5 / 9})^{11 / 5}}}}\; .
\end{equation}
Therefore for small $\delta$ this EOS describes cold NM.

From equations (\ref{EoGR}),  after integration, we have
\begin{equation}
\label{gd1} e^{ - \lambda}  = 1 - (2G / c^{2})m(r) / r, \quad m(r)
= 4\pi/c^2 {\int\limits_{0}^{r}} \varepsilon (r)r^{2}dr \; ,
\end{equation}
where $\varepsilon(r)$ is the energy density of NM and $m(r)$ is
the mass of sphere within a radius $r$. The Oppenheimer-Volkoff
for the equilibrium of NS is~\cite{Land1}
\begin{equation}
\label{Op-Volk}
 - {\frac{{dp}}{{dr}}} = {\frac{{\left( {\varepsilon + p } \right)G[m(r) +
4\pi r^{3}p(r)/ c^{2}]}}{{ rc^{2}[r - 2(G / c^{2})m(r)]}}}.
\end{equation}
To present  Eqs. (\ref{gd1})-(\ref{Op-Volk}) in dimensionless
form, we use the  radial coordinate $x$ defined as $x= r/L_{\ast}$
where the characteristic length $L_{\ast} = c / \sqrt {4\pi \rho
_{\ast}  G} = 2.8 \cdot 10^{5}\, \mathrm{cm}$. In dimensionless
units the mass within a radius $r$ is
\begin{equation}
\label{mass} \mu (x) = {\int\limits_{0}^{x} {\delta (x)x^{2}dx}}.
\end{equation}
In order to get the physical mass we multiply $\mu$ by
 $ m_{\ast} = 4\pi L_{\ast}^{3}\rho_{\ast}  = 3.8 \cdot
10^{33}\,\mathrm{g}=1.91M_{\odot}.$

Let us denote $\rho_{\ast}L_{\ast}^2=2/\kappa\, c^2$, and  if
arguments $(r,\eta)$ are measured in units of $L_{\ast}$,
dimensionless pressure ${\Pi}$ correlates with representative of
pressure $\mathit{\Pi}(\xi)$ as ${\Pi}=\mathit{\Pi}(\xi)/\eta^2$.

\begin{figure}[h]
\includegraphics[width=125mm, height =60 mm]{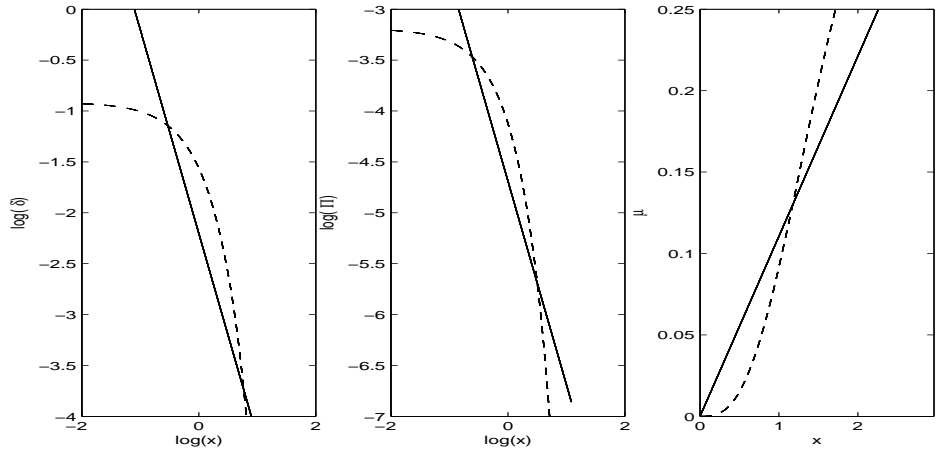}
\caption{ The profiles of the  density distribution
$\delta$,  pressure $\Pi$, and mass $\mu$ as function $x$.
 The dashed  lines are obtained for the HWW EOS
 with central density of $\delta=0.4$. The solid lines correspond
 to the analytical approximation (\ref{approx1}) with $n=1/12$. }
 \label{fig1}
\end{figure}

 Oppenheimer-Volkoff's equation (\ref{Op-Volk}) in
dimensionless variables can be presented as
\begin{equation}
\label{Op-Volk_1}
 - {\frac{{d\Pi} }{{dx}}} = {\frac{{(\delta + \Pi )(\Pi x^{3} + \mu
(x))}}{{x(x - 2\mu (x))}}}.
\end{equation}
 This
equation together with Eq. (\ref{mass}) and EOS (\ref{HW})
completely defines the structure of the neutron star.
 Given the density at the center and assuming a vanishing pressure at the star's boundary, a unique
solution to these equations can be found by numerical integration.
We denote the density at the center ($x=0$) by $\delta _{0}$.
The numerical solutions are shown as the dashed lines in the
three panels in  Fig.\ref{fig1}.
Also,  the solid lines in these panels show  solutions
obtained by assuming
\begin{equation}
\label{approx1} \delta = A / x^{2}, \qquad \Pi = n \delta \; .
\end{equation}
Thus we have:
\begin{equation}
 \Pi =  An / x^{2};\quad \mu (x) = A  x {\rm .}
\end{equation}
It is easy to see that this special solution satisfies to Eq.
(\ref{Op-Volk_1}) where the coefficient $A$ is defined by
expression:
\begin{equation}
\label{cA} A = 2n / (n^{2} + 6n + 1).
\end{equation}
At that the components of the metric $( - g^{11})$ and  $g_{00} =
e^{\nu} $ are   given by the formula
\begin{equation}
\label{cg} e^{ - \lambda}  = 1 - 2A,\qquad \nu = B(n)\ln (x /
x_{\ast}  ), \quad B(n) = {\frac{{2A(1 + n)}}{{1 - 2A}}},
\end{equation}
where $x < x_{\ast}$ (here $x_{\ast}$ is an integration constant).
It is interesting to note that  the $( - g^{11})$  is constant,
while   $g_{00}$ depends on $x$.

This solution has  a  conic singularity at
$x=0$. But the metric at this point has no singularity.
We   approximate
 density profiles of neutron stars by shifting this solution
at $x \to x + x_{0} $. Comparison of this simple analytical
solution with a numerical calculation is presented in
Fig.~\ref{fig1}.

Therefore approximating the NM distribution  with a power law
of $k=2$ as satisfactory  away from the center.

\end{document}